\documentstyle[epsf]{mn}

\newif\ifAMStwofonts

\ifoldfss
  \ifCUPmtlplainloaded \else
    \NewTextAlphabet{textbfit} {cmbxti10} {}
    \NewTextAlphabet{textbfss} {cmssbx10} {}
    \NewMathAlphabet{mathbfit} {cmbxti10} {} 
    \NewMathAlphabet{mathbfss} {cmssbx10} {} 
  \fi
  \ifAMStwofonts
    \ifCUPmtlplainloaded \else
      \NewSymbolFont{upmath} {eurm10}
      \NewSymbolFont{AMSa} {msam10}
      \NewMathSymbol{\upi}     {0}{upmath}{19}
      \NewMathSymbol{\umu}     {0}{upmath}{16}
      \NewMathSymbol{\upartial}{0}{upmath}{40}
      \NewMathSymbol{\leqslant}{3}{AMSa}{36}
      \NewMathSymbol{\geqslant}{3}{AMSa}{3E}

    \fi
  \fi
\fi 

\ifnfssone
  \newmathalphabet{\mathit}
  \addtoversion{normal}{\mathit}{cmr}{m}{it}
  \addtoversion{bold}{\mathit}{cmr}{bx}{it}
  \newmathalphabet{\mathbfit} 
  \addtoversion{normal}{\mathbfit}{cmr}{bx}{it}
  \addtoversion{bold}{\mathbfit}{cmr}{bx}{it}
  \newmathalphabet{\mathbfss} 
  \addtoversion{normal}{\mathbfss}{cmss}{bx}{n}
  \addtoversion{bold}{\mathbfss}{cmss}{bx}{n}
  \ifAMStwofonts
    \ifCUPmtlplainloaded \else
      \UseAMStwoboldmath
      \makeatletter
      \new@mathgroup\upmath@group
      \define@mathgroup\mv@normal\upmath@group{eur}{m}{n}
      \define@mathgroup\mv@bold\upmath@group{eur}{b}{n}
      \edef\UPM{\hexnumber\upmath@group}
      \new@mathgroup\amsa@group
      \define@mathgroup\mv@normal\amsa@group{msa}{m}{n}
      \define@mathgroup\mv@bold\amsa@group{msa}{m}{n}
      \edef\AMSa{\hexnumber\amsa@group}
      \makeatother
      \mathchardef\upi="0\UPM19
      \mathchardef\umu="0\UPM16
      \mathchardef\upartial="0\UPM40
      \mathchardef\leqslant="3\AMSa36
      \mathchardef\geqslant="3\AMSa3E
    \fi
  \fi
\fi 

\ifnfsstwo
  \DeclareMathAlphabet{\mathbfit}{OT1}{cmr}{bx}{it}
  \SetMathAlphabet\mathbfit{bold}{OT1}{cmr}{bx}{it}
  \DeclareMathAlphabet{\mathbfss}{OT1}{cmss}{bx}{n}
  \SetMathAlphabet\mathbfss{bold}{OT1}{cmss}{bx}{n}
  \ifAMStwofonts
    \ifCUPmtlplainloaded \else
      \DeclareSymbolFont{UPM}{U}{eur}{m}{n}
      \SetSymbolFont{UPM}{bold}{U}{eur}{b}{n}
      \DeclareSymbolFont{AMSa}{U}{msa}{m}{n}
      \DeclareMathSymbol{\upi}{0}{UPM}{"19}
      \DeclareMathSymbol{\umu}{0}{UPM}{"16}
      \DeclareMathSymbol{\upartial}{0}{UPM}{"40}
      \DeclareMathSymbol{\leqslant}{3}{AMSa}{"36}
      \DeclareMathSymbol{\geqslant}{3}{AMSa}{"3E}
    \fi
  \fi
\fi 

\ifCUPmtlplainloaded \else
  \ifAMStwofonts \else 
    \def\upi{\pi}
    \def\umu{\mu}
    \def\upartial{\partial}
  \fi
\fi

\title[Radio sources at 95\,GHz] {The extragalactic radio--source population 
at 95\,GHz }

\author[E.M. Sadler et al. ]{
\parbox[t]{\textwidth}{
Elaine M.\ Sadler$^1$, 
Roberto Ricci$^2$,
Ronald D. Ekers$^3$, 
Robert J. Sault$^4$, 
Carole A. Jackson$^3$, 
Gianfranco De Zotti$^{5,6}$ }
\vspace*{6pt} \\ 
$^1$School of Physics, University of Sydney, NSW 2006, Australia \\
$^2$Dept.\ of Physics and Astronomy, University of Calgary, 2500 University Drive NW, Calgary AB, T2N 1N4, Canada \\
$^3$Australia Telescope National Facility, CSIRO, P.O.\ Box 76, 
Epping, NSW 1710, Australia \\  
$^4$School of Physics, University of Melbourne, Victoria 3010, Australia \\
$^5$SISSA/ISAS, Via Beirut 2-4, I-34014 Trieste, Italy \\
$^6$INAF, Osservatorio Astronomico di Padova, Vicolo del'Osservatorio 5,
I-35122, Padova, Italy \\
}
 
\pagerange{\pageref{firstpage}--\pageref{lastpage}}
\pubyear{2007}

\begin{document}

\maketitle

\label{firstpage}

\begin{abstract}
We have used the Australia Telescope Compact Array (ATCA) at 95\,GHz 
to carry out continuum observations of 130 extragalactic radio sources 
selected from the Australia Telescope 20\,GHz (AT20G) survey.  
We use a triple--correlation method to measure 
simultaneous 20 and 95\,GHz flux densities for these objects, 
and over 90\% of our target sources are detected at 95\,GHz. 
We demonstrate that the ATCA can robustly measure 95\,GHz flux densities 
with an accuracy of $\sim$10\% in a few minutes for sources stronger 
than about 50\,mJy. 

We measure the distribution of radio spectral indices in a 
flux--limited sample of extragalactic sources, 
and show that the median 20--95\,GHz spectral index does not vary 
significantly with flux density for S$_{20}>150$\,mJy. This finding 
allows us to estimate the extragalactic radio source counts at 
95\,GHz by combining our observed 20--95\,GHz spectral--index distribution 
with the accurate 20\,GHz source counts measured in the AT20G survey.  

Our derived 95\,GHz source counts at flux densities above 80\,mJy 
are significantly lower than those found by several previous studies. 
The main reason is that most radio sources with flat or rising spectra 
in the frequency range 5--20\,GHz show a spectral turnover between 
20 and 95\,GHz.  
As a result, there are fewer 95\,GHz sources (by almost a factor of two 
at 0.1\,Jy) than would be predicted on the basis of extrapolation 
from the source populations seen in lower--frequency surveys. 
We also derive the predicted confusion noise in CMB surveys at 95\,GHz 
and find a value 20--30\% lower than previous estimates. 

The 95\,GHz source population at the flux levels probed by this study 
is dominated by QSOs with a median redshift $z\sim1$.  
We find a correlation between optical magnitude and 95\,GHz flux density 
which suggests that many of the brightest 95\,GHz sources may be 
relativistically beamed, with both the optical and millimetre 
continuum significantly brightened by Doppler boosting.  
\end{abstract}

\begin{keywords}
radio continuum: galaxies --- galaxies: active --- 
quasars: general --- radio continuum: general --- 
cosmic microwave background --- techniques: interferometric 
\end{keywords}

\section{Introduction}
The radio sky at 90--100\,GHz remains largely uncharted territory.  
At frequencies above 5\,GHz, traditional surveys of even modest 
areas of sky become impractical because of the small field of view 
of current telescopes. For example, the effective field of view of the NRAO VLA 
telescope is 0.22\,deg$^2$ at 1.4\,GHz, but less than 0.0002\,deg$^2$ 
at its highest frequency of 50\,GHz.  Most radio telescopes designed to 
operate at millimetre wavelengths have similarly small fields 
of view\footnote{In general, the area of sky covered by the 
primary beam of a radio telescope scales as $\nu^{-2}$ for increasing 
frequency $\nu$.}, 
so that blind surveys of large areas of sky are only possible by using 
specially-designed instruments or techniques.  

\begin{figure}
\begin{center}
\includegraphics{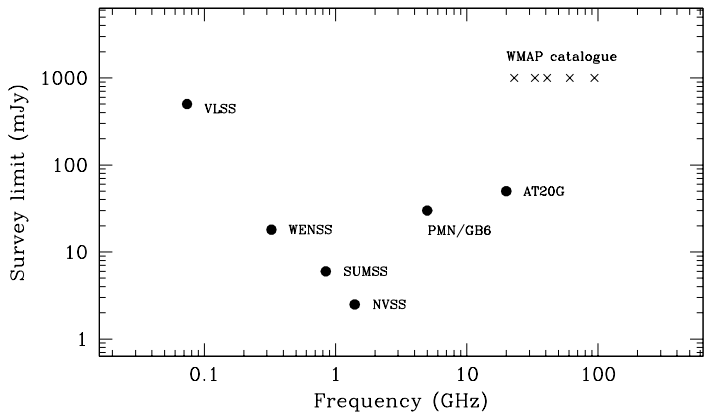} 
\vspace{6cm}
\caption{Detection limits of the most sensitive large--area radio surveys 
available at frequencies from 74\,MHz to 94\,GHz.  Only surveys which cover 
at least 25\% of the sky have been included. The surveys plotted are the VLA 
Low--Frequency Sky Survey (VLSS, Cohen et al.\ 2007), Westerbork Northern 
Sky Survey (WENSS, Rengelink et al. 1997), Sydney University Molonglo Sky Survey 
(SUMSS, Bock et al.\ 1999),NRAO VLA Sky Survey (NVSS, Condon et al.\ 1998), 
single-dish PMN and GB6 surveys (Griffith \& Wright 1993; Gregory et al.\ 1996),  
Australia Telescope 20\,GHz Survey (AT20G, Ricci et al.\ 2004) and the WMAP 
extragalactic source catalogue (Hinshaw et al.\ 2007). 
 }
\label{fig.surv}
\end{center}
\end{figure}

Figure \ref{fig.surv} summarizes the sensitivity limits of large--area radio 
continuum surveys which cover at least a quarter of the sky. 
There are currently only two large--area surveys above 5\,GHz. 
The WMAP Point Source Catalogue (Bennett et al.\ 2003; Hinshaw et al.\ 2007) 
detects sources down to 0.8--1.0\,Jy in five frequency bands from 23--94\,GHz.  
The AT20G survey (Ricci et al.\ 2004; Sadler et al.\ 2006) 
uses a wide--band (8\,GHz bandwidth) analogue correlator to scan the sky rapidly 
at 20\,GHz,  detecting sources down to a limit of $\sim40$\,mJy, and 
currently represents the highest radio frequency at which a blind survey 
of a large fraction of the sky has been carried to a detection limit 
below 100\,mJy. 

There are several reasons why we would like to learn more about 
the extragalactic radio--source population at frequencies well 
above those probed by existing large--area surveys.  
Corrections for extragalactic foreground confusion (e.g. De Zotti et 
al.\ 2005) will be critical for next--generation surveys of the 
Cosmic Microwave Background (CMB) like the Planck mission, which 
aims to measure CMB anisotropies on angular scales of 5--30\,arcmin 
in the frequency range 30--350\,GHz.  Such corrections require an accurate 
knowledge of the foreground point--source population at frequencies of 
30\,GHz and above. 

Observations at 95\,GHz are particularly relevant to Planck, since the 
Planck 100\,GHz channel and the nearby 143\,GHz channel are the ``cleanest'' 
ones for CMB studies in both temperature and polarization on small angular 
scales, where the main foreground contamination comes from extragalactic 
sources. The 100 and 143\,GHz channels will also be the most sensitive 
Planck channels\footnote{Although the quoted Planck sensitivity 
($\Delta$T/T per pixel) is slightly better at 30\,GHz than at 100 
and 143\,GHz, the smaller pixel size (in arcmin) at high frequencies 
means that the 100 and 143\ GHz channels are far more sensitive when scaled 
to the same angular resolution.}\ 
(Lamarre et al.\ 2003), far exceeding the performances 
of WMAP, so that an accurate subtraction of point source contamination 
at these frequencies will be very important. 

A second motivation arises because current models of radio--galaxy evolution 
imply that very young radio sources (tens to hundreds of years old) should have radio 
spectra which peak above 5\,GHz (the so--called Gigahertz--Peaked Spectrum, 
or GPS sources; O'Dea 1998), with the spectra of younger sources peaking 
at progressively higher frequencies.  Such objects are rare, since they 
represent a very short--lived evolutionary stage, but are predicted to be 
extremely luminous and hence easily detectable out to high redshift in 
high--frequency radio surveys.  Until now, models for the luminosity evolution 
of young radio galaxies have been almost totally unconstrained by the 
available data; and the need for a large, homogeneously--selected 
sample of GPS sources is well-recognized (e.g. Snellen et al.\ 2000). 
An analysis of evolutionary properties of GPS galaxies has recently been
carried out by Tinti \& De Zotti (2006). 

In addition to these science goals, 95\,GHz observations are relevant to 
the calibration strategy for the Atacama Large Millimetre Array 
(ALMA; De Breuck 2005), which will need to do rapid 
switching between object and calibrators to achieve the required phase stability.
This ideally requires a network of calibrators no more than 1--2 degrees apart and stronger 
than 50-100~mJy at 90~GHz. The existing proposal (Holdaway et al.\ 2004) is to find 90~GHz
calibrators by selecting candidates from low-frequency (1--5~GHz) surveys and using ALMA 
itself to measure their 90~GHz flux density. We have already suggested (Sadler et al.\ 2006) 
that a more efficient alternative would be to select ALMA calibrators from the high-frequency 
Australia Telescope 20\,GHz (AT20G survey).  One aim of the present study is therefore 
to measure the surface density of AT20G sources at 90--100\,GHz.  

Early work in this area dates back to the 1970s.  Owen et al.\ (1978) used the 
NRAO 11\,m radio telescope at Kitt Peak to measure the 90\,GHz flux density of 
237 strong sources which were known to have flat or inverted radio spectra at 
centimetre wavelengths. They also made near-simultaneous observations of about 
100 of these sources at 1.4, 4.6, 15 and 22\,GHz using dishes of the NRAO VLA.  
Their main findings were that most of their sources had flat spectra ($\alpha\sim0$) 
over the whole frequency range 1--90\,GHz, and that the level of radio variability 
at 90\,GHz was lower than expected (most sources varied by less than 30\% on a 
one--year timescale).  Owen, Helfand \& Spangler (1981) observed 25 strong 
extragalactic millimetre sources at $\sim1$\,keV  with the {\it Einstein}\ X--ray 
obsetvatory, and found a strong correlation between their X--ray and 90\,GHz 
flux densities.  They noted that this is consistent with a synchrotron 
self--Compton model in which the X--ray emission arises from the Compton 
scattering of millimetre radio photons off the relativistic electrons which 
created them. 

Steppe et al.\ (1988) observed a sample of 294 strong ($>1$\,Jy at 5\,GHz) 
extragalactic sources at 90\,GHz with the IRAM 30\,m telescope, and also 
measured  20--90\,GHz spectral indices for a subset of these objects.  
They found that only about 5\% of 
compact, flat--spectrum radio sources had inverted ($\alpha>0.1$) spectra at 
20--90\,GHz, and that most of these were variable sources undergoing an outburst. 

More recent continuum work at 90--100\,GHz has mainly focussed on variability 
studies of targeted samples of objects such as QSOs and blazars 
(e.g. Landau et al.\ 1980; Ter\"asranta et al.\ 2004; Hovatta et al.\ 2007).  
All these studies have been based on sources pre--selected at frequencies 
of 5\,GHz or below.  The WMAP catalogue (Hinshaw et al.\ 2007) provides 
the first complete sample of radio sources selected in the 90\,GHz band, 
but only 121 of the 323 discrete sources listed in the 3--year catalogue 
are detected at 94\,GHz (limiting flux density 0.8\,Jy).  

Throughout this paper, we use $H_0$ = 70 km s$^{-1}$ 
Mpc$^{-1}$, $\Omega_m$ = 0.3 and  $\Omega_\Lambda$ = 0.7.

\section{ATCA 95\,GHz observations }
The data analysed in this paper were obtained at the Australia Telescope Compact Array 
(ATCA) in two observing runs during 2005 and 2006 (see Table \ref{tab.obs} for 
details of the observations). 

\begin{table}
\begin{center}
\begin{tabular}{lccc}
\hline
\multicolumn{1}{c}{Date} & \multicolumn{1}{c}{ATCA} & 
\multicolumn{1}{c}{$\nu_{\rm cen}$} & \multicolumn{1}{c}{Notes} \\
\hline
2005 Jul 1--3   & H75A & 18.8, 21.1, 93.5, 95.6 & 2$\times$24\,hr \\
2006 Sep 16--19 & H75A & 18.8, 21.1, 93.5, 95.6 & 3$\times$12\,hr \\
\hline
\end{tabular}
\caption{Log of our ATCA observations at 95\,GHz. }
\label{tab.obs}
\end{center}
\end{table}

The ATCA was used in mosaic mode to minimize the slewing time between 
sources, with the target sources observed in groups of 5--7 objects 
which were close (within about 10\,degrees on the sky) 
in both right ascension and declination. 

\subsection{Sample selection}
All our target sources were selected from the Australia Telescope 
20\,GHz (AT20G) survey of the southern sky (Ricci et al.\ 2004; 
Sadler et al.\ 2006) as noted in \S1. The target selection 
criteria were different in 2005 and 2006, and are summarized in Table 
\ref{tab.sel}.  

An important feature of the AT20G survey is that sources selected at 
20\,GHz are also observed near--simultaneously at 5 and 8\,GHz, so 
that radio spectral indices measured in the 5--20\,GHz frequency range 
are not affected by variability.  A recent description of the AT20G 
data pipeline is given by Massardi et al.\ (2008). 

\begin{table}
\begin{center}
\begin{tabular}{lrlll}
\hline
\multicolumn{1}{c}{Run} & \multicolumn{1}{c}{N}& \multicolumn{1}{c}{20\,GHz flux} & 
\multicolumn{1}{c}{Spectral} & \multicolumn{1}{c}{Notes} \\
  &  &  \multicolumn{1}{c}{density} & \multicolumn{1}{c}{index cut} & \\
\hline
2005  & 59     & $>50$\,mJy  & $\alpha_8^{20}>0$ & `Inverted-spectrum' \\
2005  &  9     & $>300$\,mJy & ... & AT calibrator sample \\
2006  & 70     & $>150$\,mJy & ... & Flux--limited sample \\
\hline
\end{tabular}
\caption{Sample selection for the 2005 and 2006 runs. All targets were also chosen to 
lie in the declination range $-30^\circ$ to $-50^\circ$ to ensure that they could be 
observed at low airmass.  }
\label{tab.sel}
\end{center}
\end{table}

In July 2005 we observed a sample of 59 sources selected from the AT20G survey to have 
inverted (rising) radio spectra between 8 and 20\,GHz. Nine strong AT20G sources which 
are ATCA calibrators were also observed to provide a consistency check on our final 
flux--density scale at 95\,GHz. All these sources were also chosen to lie at declinations 
between $-30^\circ$ and $-50^\circ$ to ensure that they could be observed at low airmass. 
Our aim was to see how many of the `inverted--spectrum' AT20G sources showed a high--frequency 
turnover above 20\,GHz and so were candidates for extreme GPS sources. 

In September 2006, we observed a flux--limited (S$_{20}>150$\,mJy) sample of 70 AT20G sources, 
again between declination $-30^\circ$ and $-50^\circ$.  Our aim here was to use the 
distribution of 20--95\,GHz radio spectral indices to estimate the overall surface density 
of extragalactic radio sources at 95\,GHz. Eight sources belonged to both the 2005 and 2006 
samples, giving a total of 130 individual sources with 95\,GHz measurements in the two--year 
program. 

\subsection{Antenna configuration and observing frequencies}
We used the ATCA in its most compact configuration, the hybrid H75A array, which has a maximum 
baseline of 75\,m and is designed to provide good $uv$ coverage and phase stability for short 
observations at millimetre frequencies.  The ATCA field of view is roughly 2.5\,arcmin 
at 20\,GHz and 30\,arcsec at 95\,GHz, and the angular resolution in the H75 array is around 
40\,arcsec at 20\,GHz and 10\, arcsec at 95\,GHz.  

Since one of our goals was to measure reliable 
20--95\,GHz spectral indices for the AT20G sources, and some AT20G sources are known 
to be variable at 20\,GHz (Sadler et al.\ 2006), we observed all our target sources at both 
20 and 95\,GHz in the 2005 and 2006 runs. 
The observing frequencies (IFs) listed in Table \ref{tab.obs} were selected to lie in regions  
where the ATCA receivers have their best Tsys performance. 

\subsection{Calibration techniques}
At the time of these observations the ATCA 3\,mm system did not have a noise 
diode, so to place visibility amplitudes on 
a temperature scale a vane (paddle) needed 
to be interposed in front of the receiver in the antenna 
vertex room at regular intervals for Tsys calibration. 
This was done at the start of each observing session, and also 
repeated just before starting the observation of each source group 
(at intervals of about half an hour).  

\subsubsection{Flux--density calibration} 
At 20\,GHz, the absolute flux--density scale was set using the standard ATCA primary calibrator 
PKS\,1934$-$638. At 95\,GHz, however, PKS\,1934$-$638 is far too weak to provide a reliable 
calibration of the flux--density scale.  We therefore followed the usual practice and used 
the planet Uranus as our flux calibrator at 95\,GHz.  Although the angular size of this planet 
is significantly smaller than the ATCA beam, it cannot be assumed to be a point source.  
A model of its intensity as a function of frequency and baseline length is therefore used to 
perform the calibration.   To minimize any gain-elevation effects (i.e. possible variations 
in antenna efficiency as a function of elevation) on the primary calibration, 
we always observed the planet close to transit and followed or preceded by a 
target source group at a similar elevation. 

\subsubsection{Antenna pointing calibration} 
\begin{figure}
\begin{center}
\includegraphics{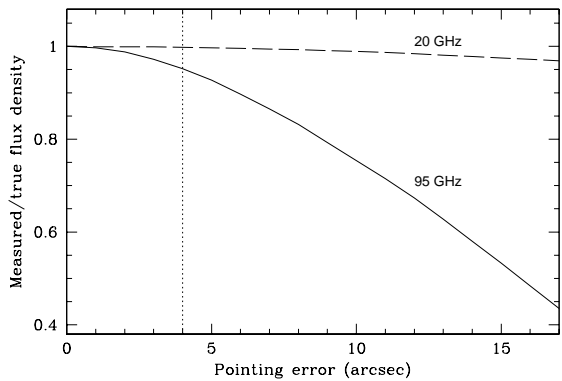} 
\vspace{6cm}
\caption{The effect of antenna pointing errors on the measured flux density 
at 20 and 95\,GHz. 
For our 2005 and 2006 runs, pointing and tracking errors are expected to 
be less than 4.0\,arcsec (shown by the dotted vertical line), except on rare 
occasions when the wind is both strong and gusty. 
Pointing and tracking errors should therefore have at most a small effect 
on the measured flux densities and 20--95\,GHz spectral index distribution. }
\label{fig.pb}
\end{center}
\end{figure}

The ATCA primary beam at 95\,GHz is only 30\,arcsec FWHM, and the pointing and tracking 
requirements for operation at this high frequency have been discussed in detail by 
Kesteven (1998).  The effects of a pointing offset can be seen in Figure \ref{fig.pb}.  
The main factors affecting the positional accuracy of the ATCA are pointing and tracking 
errors (typically 1.5\,arcsec rms on each axis when reference pointing is used), atmospheric 
fluctuations (up to 3.0\,arcsec rms for sources observed close to the zenith), and wind 
loading (which starts to have a significant effect when the wind speed rises above 
10--15 km\,hr$^{-1}$, especially when conditions are gusty).  Also contributing to the 
error budget is the uncertainty in the individual AT20G source positions (typically less than 
1.0\,arcsec).  

The uncorrected ATCA antenna pointing 
errors in typical observing conditions can be up to 20\,arcsec.  It is therefore essential 
to use reference pointing when observing at high frequencies.  A reference pointing check 
using a strong phase calibrator was therefore done whenever moving 
to a new source group.  This allowed a new pointing solution to be calculated and applied to 
the sources in each group, reducing the final pointing errors to a few arcsec or less.  
We generally did our reference pointing calibrations at 20\,GHz rather than 95\,GHz, because 
the higher signal-to-noise ratio and larger field of view at 20\,GHz 
allows a more accurate solution. 

\subsection{2005 observing run}
Both the effective detection limit for continuum sources at 95--100\,GHz 
and the accuracy of the derived flux densities are strongly affected by 
atmospheric transparency, phase stability and decorrelation, and the 
general performance of the ATCA 3\,mm system for weak 
continuum sources was unknown at the time of our 2005 run.  
We therefore took a conservative approach and repeated the same set of 
observations on each of the two days we were allocated.  

Since all our target sources were known (from AT20G images at 20\,GHz) 
to be point sources at high frequency, we chose to measure flux densities 
using a triple correlation technique rather than imaging our target fields.  
The triple correlation technique is discussed briefly in \S3 of this paper, 
and in more detail in a companion paper by Ricci, Sault \& Ekers (in preparation). 

In the 2005 run, each group of target sources was observed three times at 
different hour angles to provide a range in $uv$ coverage.  The total integration 
time for each source at 95\,GHz was $3\times4$\,minutes on each day.  On the second day, 
each group of sources was observed at both 95\,GHz and 20\,GHz, with the total observing 
time at 20\,GHz being $3\times1$\,minute.  To minimize the effects of airmass, all 
observations were made at hour angles less than $\pm1.5$ hours from the zenith. 

\subsection{2006 observing run}
During our 2005 run, we found that observing conditions at 95\,GHz were significantly better 
at night than during the day.  This appears to be generally the case at the ATCA, even 
during the winter months, as discussed recently by Middelberg et al.\ (2006).   
We therefore chose to carry out our 2006 observing campaign in three consecutive 12\,hr 
nighttime slots.  The array configuration and frequency set--up were the same as in 2005.  

Our 2006 sample of 70 AT20G sources was split into nine groups which were observed 
at 20 and 95\,GHz on either the first or the second nights. Each source was observed 
between two and four times, with the integration time for each observation being 80\,s 
at 95\,GHz and 40\,s at 20\,GHz.  The integration times were shorter than in 2005 because 
most of the 2006 sources were significantly brighter at 20\,GHz than those observed in 2005. 
On the third night, we reobserved 32 sources which did not appear to show 
an obvious 95\,GHz detection on the first two days.  These objects were re-observed for 
$3\time3$\,min cuts at 95\,GHz. 
 
\section{Data reduction}
The raw visibility data were reduced using the astronomical software package MIRIAD 
(Sault et al.\ 1995). The 95\,GHz and 20\,GHz bandpasses were calibrated using the 
strong ATCA high--frequency calibrator PKS\,1921$-$293.

\subsection{Flux--density measurements} 
Atmospheric turbulence creates strong phase instability in the 95\,GHz band and 
target sources and phase calibrators need to be switched very frequently 
(i.e. on timescales of a few minutes or less) to avoid the loss of total 
intensity flux through phase decorrelation.  

Because of the large number of sources we wanted to observe in both 
2005 and 2006, it would have been difficult to spend enough time on 
phase calibrators to correct for phase decorrelation in the standard way.  
We therefore chose not to observe any secondary (phase) calibrators in 
our 2006 run.  Instead, we chose to reduce our 2005 and 2006 data 
using the triple correlation (phase closure) method as noted earlier.  

The triple correlation method has also been used by Sault et al.\ when monitoring 
flux densities of ATCA calibrators at 3\,mm, and is robust against phase 
variations during the target observation.  It is described in detail by Cornwell (1987) 
and Thompson et al.\ (2001).  The application of this technique to ATCA millimetre data 
is discussed by Ricci et al.\ (in preparation), who have also used a 
set of Monte Carlo simulations to test the robustness of this technique for 
the detectability and flux density measurements of the 95~GHz data presented in this paper. 

Table \ref{tab_err} lists the estimated flux--density errors for our 2005 and 2006 data. 
These were calculated from the scatter in the triple--correlation amplitudes by taking 
different scans and different spectral windows (IFs) of the same source separately. 
At 20\,GHz, the flux density errors are significantly higher in 2006 than in 2005.  
This is probably because the 2005 20\,GHz data were taken in very good observing conditions, 
whereas the 2006 data at the same frequency were much noisier.

\begin{table}
\begin{center}
\begin{tabular}{lcc}
\hline
 & \multicolumn{2}{c}{Flux--density error estimates } \\
  & \multicolumn{1}{c}{20\,GHz} & \multicolumn{1}{c}{95\,GHz} \\
\hline
2005  &  5.4\%    & 11.8\% \\
2006  & 10.7\%    & 10.2\%  \\
\hline
\end{tabular}
\caption{Error estimates for the flux densities measured in 2005 and 2006, derived 
using the methods of Ricci et al.\ (2007).  }
\label{tab_err}
\end{center}
\end{table}

\subsection{Data tables}
Tables 4 and 5 list the individual sources observed at 20 and 95\,GHz in our 
2005 and 2006 runs. The main information contained in these tables is as follows: 
\begin{enumerate}
\item
The AT20G working name, used to track individual sources through the data pipeline. 
A letter `C' after the name indicates that the source is also an ATCA calibrator. 
\item
The J2000 radio position of the source, measured from AT20G 20\,GHz images. 
\item
Alternate source name, where available. 
\item
Blue (B$_{\rm J}$) magnitude of the optical counterpart if any, from the 
Supercosmos catalogue (Hambly et al.\ 2001).  As discussed in \S4.7, an optical object 
is accepted as the radio ID if it lies within 2.5\,arcsec of the AT20G radio position. 
\item
Supercosmos classification of the optical ID, where 1 indicates a galaxy 
and 2 a stellar object (candidate QSO). 
\item 
Simultaneous ATCA 95\,GHz and 20\,GHz flux densities measured in 2005/6 in this project. 
\item
Near--simultaneous ATCA 20, 8 and 5\,GHz flux densities measured by the AT20G team in 
November 2004.  The listed values are from the July 2007 run of the AT20G data pipeline. 
\item
Lower--frequency flux--density measurements from the 1.4\,GHz NVSS (Condon et al.\ 1998) 
and 843\,MHz SUMSS (Mauch et al.\ 2003) catalogues.  The SUMSS values are taken from 
v2.0 of the catalogue (released August 2007). 
\item
The spectroscopic redshift of the optical counterpart, where this is available either from the 
NASA Extragalactic Database (NED) or from the 6dFGS Third Data Release (Jones et al., in preparation).  
\item 
NED classification as QSO (Q) or galaxy (G), where available. These classifications are 
taken from papers in the literature and usually (but not always) agree with the Supercosmos 
classification. 
\end{enumerate}

Appendix A contains some notes on the properties of individual sources. 

\section{Data Analysis}

\subsection{Stability of the 95\,GHz flux--density measurements}
\begin{figure}
\begin{center}
\includegraphics{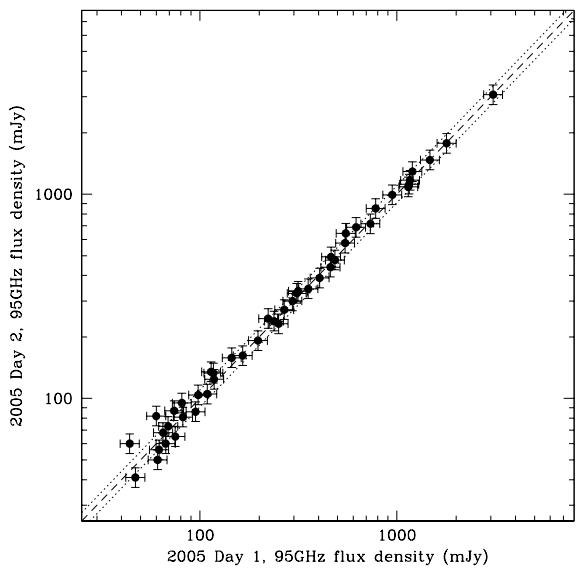} 
\vspace{8cm}
\caption{Comparison of two independent 
95\,GHz measurements (taken one day apart) of AT20G sources  
observed in our July 2005 run. Dotted lines show a 10\% deviation 
from equality. }
\label{fig.comp}
\end{center}
\end{figure}

A comparison of the 95\,GHz flux densities measured on the first and 
second days of our 2005 run is shown in Figure \ref{fig.comp}. 
The flux densities measured on the two days agree to within 10\% 
(i.e. within the quoted error bars from Table \ref{tab_err}), 
even though conditions on the first day were cloudy with poor phase 
stability. We therefore conclude that our techniques for measuring 
95\,GHz flux densities yield results which are stable and reproducible. 
This is important, because we are measuring continuum sources which are 
significantly weaker than have previously been observed with the ATCA in the 
95\,GHz band.

\setcounter{table}{5}\begin{table*}
\begin{minipage}{170mm} 
\begin{center}
\begin{tabular}{@{}lrrrrrrrrrrrrrr@{}}
\hline
\multicolumn{1}{c}{Name} & \multicolumn{6}{c}{ -------- S20 (mJy) ---------- }& \multicolumn{4}{c}{ --- S95 (mJy) -----} & 
\multicolumn{4}{c}{ ----- Spectral index $\alpha^{20}_{95}$ -----} \\
  &  2004 & $\pm$ & 2005 & $\pm$ & 2006 & $\pm$ & 2005 & $\pm$ & 2006 & $\pm$ & 2005 & $\pm$ & 2006 & $\pm$ \\
\hline
0013$-$3954  & 1609 & 33 & 1420 & 77 &  984 &105 &  716 & 84 &  496 & 51 & $-$0.44 & 0.12 & $-$0.44 & 0.14 \\
0106$-$4034  & 2146 & 79 & 2632 &142 & 4160 &445 & 1775 &209 & 2199 &224 & $-$0.25 & 0.12 & $-$0.41 & 0.14 \\ 
0136$-$4044  &  350 & 11 &  380 & 21 &  274 & 29 &  344 & 41 &  190 & 19 & $-$0.06 & 0.12 & $-$0.23 & 0.14 \\ 
0421$-$4817  &  452 & 13 &  513 & 28 &  321 & 34 &  300 & 35 &  212 & 22 & $-$0.34 & 0.12 & $-$0.27 & 0.14 \\ 
2056$-$4714  & 1711 & 55 & 1993 &108 & 3246 &347 & 1086 &128 & 2453 &250 & $-$0.39 & 0.12 & $-$0.18 & 0.14 \\  
2151$-$3027  & 1846 & 58 & 2108 &114 & 1402 &150 &  688 & 81 &  849 & 87 & $-$0.72 & 0.12 & $-$0.32 & 0.14 \\
2158$-$3013  &  350 & 10 &  388 & 21 &  844 & 90 &  327 & 39 &  903 & 92 & $-$0.11 & 0.12 & $+$0.04 & 0.14 \\ 
2235$-$4835  & 1994 & 75 &  ... & .. & 1775 &190 & 2605 &307 & 1076 &110 &   ...   & ..   & $-$0.32 & 0.14 \\
\hline
\end{tabular}
\caption{Flux--density and spectral--index measurements for the eight sources from Tables 4 and 5 which were 
observed in both the 2005 and 2006 runs. 2235$-$4835 was not observed at 20\,GHz in 2005.   }
\label{tab.repeat}
\end{center}
\end{minipage}
\end{table*}

\subsection{The inverted--spectrum and flux--limited samples} 
As noted in Table \ref{tab.sel}, our 2005 observations included 
59 inverted--spectrum sources which were selected to have rising radio 
spectral index ($\alpha>0$) between 8 and 20\,GHz.  The sample selection 
was done in July 2005, using preliminary flux--density measurements from a 
reduction of the AT20G data carried out in late 2004.  

After our 2005 observations were completed, the AT20G team 
implemented a new data pipeline which included an improved 
calibration process and a stringent set of checks designed to 
identify poor-quality data.  As a result of these 
checks, 13 galaxies in our 2005 sample (all in the RA range 17--22\,h) 
had their 8 or 20\,GHz AT20G flux measurements flagged as unreliable. 
Since we could no longer be sure that these sources qualified 
as `inverted spectrum', they were not included in our analysis. 
Two other sources no longer had rising 8--20\,GHz spectra after 
recalibration, and were also removed. 

Finally, we tightened the selection criteria for the inverted 
spectrum sample to require that the sources had $\alpha>0$ 
at 5--8\,GHz as well as 8--20\,GHz.  The aim was to restrict 
the sample to objects whose radio spectra were genuinely 
inverted over the full 5--20\,GHz AT20G frequency range, and 
to avoid contamination by sources whose spectra were 
flat ($\alpha\sim0$) at 5--20\,GHz but which happened to have 
a slightly higher flux measurement at 20\,GHz than at 8\,GHz.
A further seven (mostly weak) sources were removed by this 
test. 
All 59 sources observed in 2005 are listed in Table 4, with  
the 39 sources included in our final inverted--spectrum analysis
clearly identified. 

Table 5 lists the 70 sources in our 2006 `flux--limited' sample, 
which were selected to have 20\,GHz flux densities above 150\,mJy. 
One object (2257$-$3627, the nearby galaxy IC\,1459) 
later had its 2004 AT20G flux density at 20\,GHz flagged as unreliable. 
Although it is listed in Table 5, it was not included in our analysis.  
Our final flux--limited sample therefore contains 69 objects. 

\begin{table}
\begin{center}
\begin{tabular}{lclrc}
\hline
\multicolumn{1}{c}{Data set} & \multicolumn{1}{c}{Freq} & \multicolumn{1}{c}{Time interval} & 
\multicolumn{1}{c}{N} & \multicolumn{1}{c}{Median var.} \\
  & \multicolumn{1}{c}{GHz} &  &  & \multicolumn{1}{c}{index (\%)} \\
\hline
AT20G Pilot   & 20 & 2003 -- 2004  & 108 & 6.9 \\
Inverted-spec & 20 & 2004 -- 2005  &  37 & 6.4 \\
Flux--lim.    & 20 & 2004 -- 2006  &  69 & 6.9 \\
\hline
05--06 repeats& 20 & 2004 -- 2006  &   8 & 20.1 \\
05--06 repeats& 20 & 2005 -- 2006  &   7 & 20.3 \\
\hline
05--06 repeats& 95 & 2005 -- 2006  &   8 & 20.2 \\
\hline
\end{tabular}
\caption{Median variability index at 20 and 95\,GHz for several sets 
of data observed with the ATCA. The AT20G Pilot results for 2003--4 are taken 
from Sadler et al.\ (2006). The small sample of `repeat' sources which we 
observed in both 2005 and 2006 have a significantly higher median variability 
index than either of the parent samples from which they came.  
 }
\label{tab.var}
\end{center}
\end{table}

\subsection{Variability at 20 and 95\,GHz} 
Sadler et al.\ (2006) examined the variability of a flux--limited 
(S$_{20}>100$\,mJy) sample of 108 AT20G sources by comparing measurements 
from 2003 and 2004. They found that the general 
level of variability at 20\,GHz was quite low, with a median variability 
index for their sources of 6.9\%.  Only five sources varied by 
more than 30\% over the one--year interval. 

We can also use the 2004 AT20G measurements as a benchmark to measure 
the variability of the sources in Tables 4 and 5 at 20\,GHz.  
To do this, we calculated a debiased variability index for each object 
as described in \S6 of Sadler et al.\ (2006). 
The results are summarized in Table \ref{tab.var}.  
For both the `inverted--spectrum' sample observed in 2005 (Table 4) and 
the flux--limited sample observed in 2006 (Table 5), the median variability 
index at 20\,GHz is close to that found by Sadler et al.\ (2006) for their 
flux--limited source sample. 

In contrast to the results at 20\,GHz, we have very little data so far on variability at 
95\,GHz. There are only eight sources in common between our 2005 and 2006 data sets 
(see Table \ref{tab.repeat}), and these turn out by chance to be drawn mainly from the 
more variable objects in their parent samples.  Any results based on this small sample 
are no more than suggestive, but we see no evidence at this stage that our sources 
are more variable at 95\,GHz than at 20\,GHz (see Table \ref{tab.var}). 
The spectral index values in Table \ref{tab.repeat} also suggest that the 
measured 20--95\,GHz spectral index may not change significantly 
when a sources varies in flux density at these frequencies, though the 
size of the error bars makes it difficult to draw any strong conclusion 
from a sample of only seven repeat observations.

\subsection{Comparison with the WMAP 3--year Source Catalogue} 
Hinshaw et al.\ (2007) have recently released a catalogue of 323 radio sources detected by 
the WMAP satellite (Bennett et al.\ 2003) in five frequency bands at 23, 33, 41, 61 and 
94\,GHz. The WMAP catalogue identifies bright sources independent of their presence in other 
surveys at lower frequency.  When a source is identified with $>5\sigma$ confidence in any 
of the five bands, the flux densities in other bands are also listed if they are $>2\sigma$. 
Only 121 of the 323 catalogued WMAP sources have flux densities listed at 94\,GHz, and almost 
all of these are 23\,GHz detections. 

\begin{figure}
\begin{center}
\includegraphics{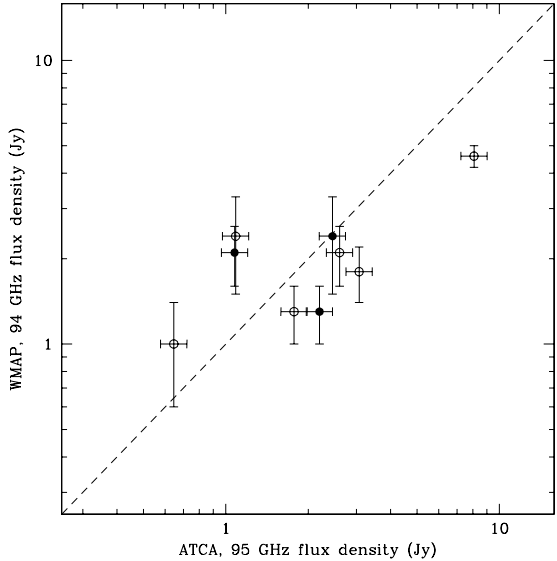} 
\vspace{8cm}
\caption{Comparison of our measured ATCA 95\,GHz flux densities with those 
listed in the WMAP 3-year source catalogue (Hinshaw et al.\ 2007) for sources 
in common (see also Table \ref{tab.wmap}).  Filled and open circles show 
data from the 2005 and 2006 ATCA runs respectively. }
\label{fig.wmap}
\end{center}
\end{figure}

Table \ref{tab.wmap} lists 
the thirteen sources in Tables 4 and 5 which were also detected by WMAP at 94\,GHz. These objects are 
also plotted in Figure \ref{fig.wmap}. The mean flux--density ratio (S$_{\rm ATCA}$/S$_{\rm WMAP}$) 
for the objects in Table \ref{tab.wmap} is $1.18\pm0.14$ at 20/23\,GHz and $1.16\pm0.06$ at 
95/94\,GHz.  Although the comparison sample is small and the sources were measured at different 
epochs, the results strongly suggest that the ATCA and WMAP data are on a common flux--density 
scale to within about 15\% at 95\,GHz. This also appears to confirm that our 95\,GHz ATCA flux 
measurements have not been strongly affected by pointing and tracking effects (see \S2.3.2).  

\begin{table*}
\begin{minipage}{170mm} 
\begin{center}
\begin{tabular}{@{}llrrrrrrrrrrrrrrrr@{}}
\hline
             &                 & \multicolumn{4}{c}{------ 2006 ------}    & \multicolumn{4}{c}{------ 2005 ------} & 
	     \multicolumn{2}{c}{--- 2004 ---} & \multicolumn{4}{c}{------- WMAP -------} \\
  AT20G name & Alt. name       & S95  & $\pm$ & S20  & $\pm$ & S95 & $\pm$ & S20& $\pm$ & S20 & $\pm$ &  S94 &$\pm$ & S23 & $\pm$ \\
             &                 & Jy   &       & Jy   &       & Jy  &       & Jy   &     & Jy   &      &  Jy       &      & Jy     &  \\ 
\hline
 0012$-$3954   &  WMAP\,202       & 0.50 & 0.05 & 0.98 & 0.10 & 0.72 & 0.08 & 1.42 & 0.08 & 1.61 & 0.03 & ... & ... & 1.4 & 0.06 \\ 
 0026$-$3512   &  PMNJ0026$-$3512 &  ... & ...  & ...  & ...  & 1.12 & 0.13 & 1.27 & 0.07 & 1.12 & 0.02 & ... & ... & 1.0 & 0.08 \\  
 0106$-$4034   &  WMAP\,171       & 2.20 & 0.22 & 4.16 & 0.45 & 1.78 & 0.21 & 2.63 & 0.14 & 2.15 & 0.08 & 1.3 & 0.3 & 1.8 & 0.05 \\
 0334$-$4008   &  WMAP\,146       & 0.35 & 0.04 & 0.55 & 0.06 & ...  & ...  & ...  & ...  & 1.27 & 0.04 & ... & ... & 1.6 & 0.06 \\
 0440$-$4333   &  WMAP\,147       & ...  & ...  & ...  & ...  & 0.44 & 0.04 & 1.53 & 0.16 & 1.95 & 0.05 & ... & ... & 3.0 & 0.07 \\
 0455$-$4615   &  WMAP\,151       &  ... & ...  & ...  & ...  & 3.07 & 0.36 & 4.42 & 0.24 & 4.16 & 0.12 & 1.8 & 0.4 & 3.7 & 0.06 \\
 0538$-$4405   &  WMAP\,148       &  ... & ...  & ...  & ...  & 8.08 & 0.95 & 7.33 & 0.40 & 5.29 & 0.22 & 4.6 & 0.4 & 5.4 & 0.06 \\
 1102$-$4404   &  PKS1059$-$438   &  ... & ...  & ...  & ...  & 0.49 & 0.06 & 0.49 & 0.03 & 0.77 & 0.02 & ... & ... & 0.6 & 0.05 \\
 1107$-$4449   &  WMAP\,166       &  ... & ...  & ...  & ...  & 0.65 & 0.08 & ...  & ...  & 1.67 & 0.04 & 1.0 & 0.4 & 1.6 & 0.05 \\
 1427$-$3305   &  WMAP\,193       &  ... & ...  & ...  & ...  & 1.47 & 0.17 & 0.87 & 0.05 & 0.77 & 0.02 & ... & ... & 3.2 & 0.07 \\
 1937$-$3958   &  PKS1933$-$40    &  ... & ...  & ...  & ...  & 1.29 & 0.15 & 1.67 & 0.09 & 1.76 & 0.05 & ... & ... & 0.7 & 0.08 \\
 2056$-$4714   &  WMAP\,208       & 2.45 & 0.25 & 3.25 & 0.35 & 1.09 & 0.13 & 1.99 & 0.11 & 1.71 & 0.06 & 2.4 & 0.9 & 1.7 & 0.06 \\
 2235$-$4835   &  WMAP\,206       & 1.08 & 0.11 & 1.78 & 0.19 & 2.61 & 0.31 & ...  & ...  & 1.99 & 0.08 & 2.1 & 0.5 & 1.7 & 0.06 \\
\hline
\end{tabular}
\caption{Comparison of flux densities for sources in common with the WMAP 3--year catalogue (Hinshaw et al.\ 2007). 
The WMAP data were taken over the period 2001 Aug to 2004 Aug inclusive and so pre-date the ATCA observations, 
which were made in 2004 Oct, 2005 Jul and 2006 Sep. }
\label{tab.wmap}
\end{center}
\end{minipage}
\end{table*}

\subsection{The 20--95\,GHz spectral--index distribution} 
\begin{figure}
\begin{center}
\includegraphics{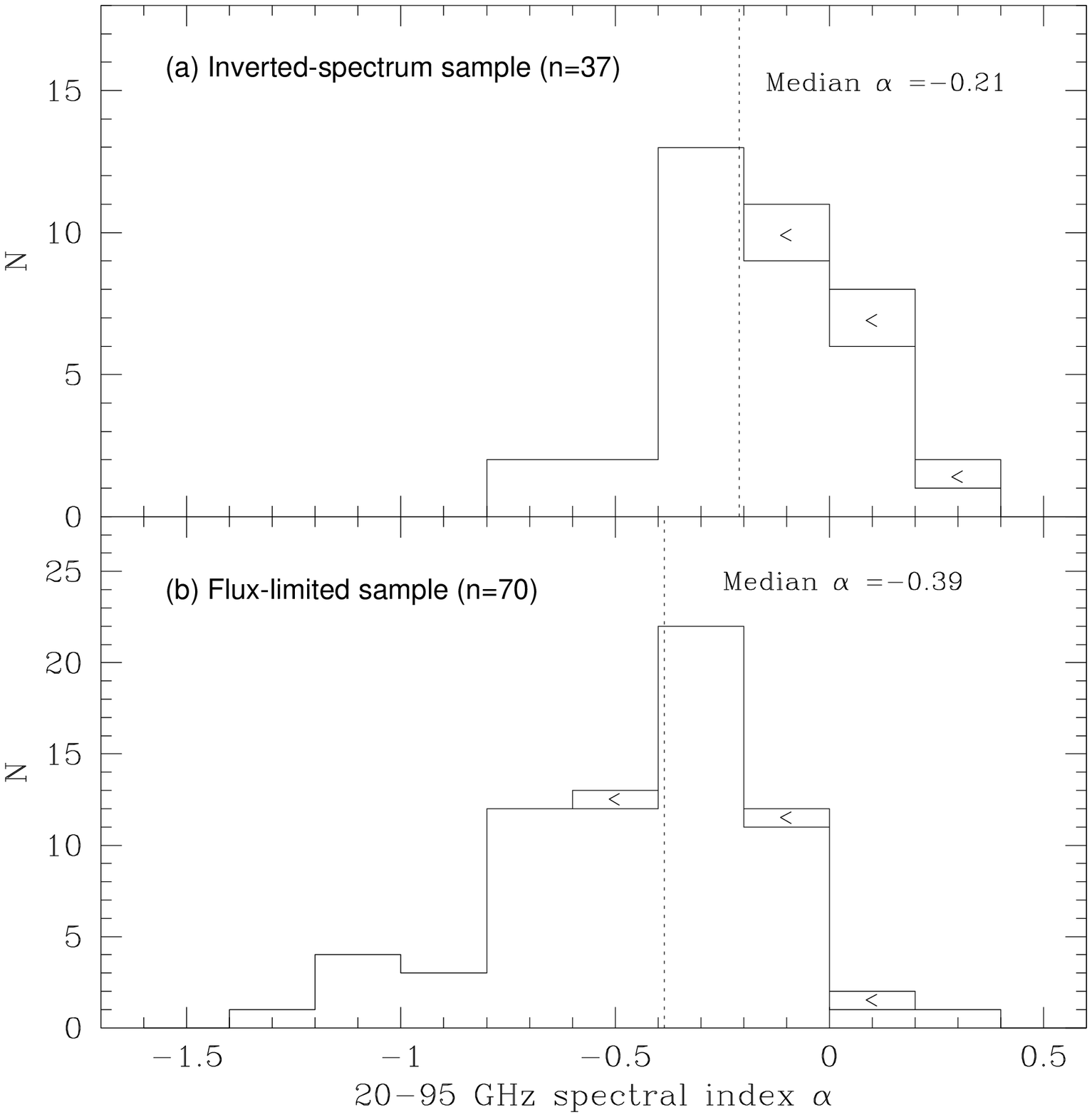} 
\vspace{8cm}
\caption{Histogram of 20--95\,GHz spectral index $\alpha$ 
for (a) the inverted--spectrum sample observed in 2005 and (b) the flux--limited 
sample observed in 2006.}
\label{fig.hist}
\end{center}
\end{figure}

As a first step in investigating the high--frequency properties of our sample,  
we calculated the two--point radio spectral index $\alpha_{20}^{95}$ for each of 
the objects in Tables 4 and 5.  For the small number of 20\,GHz sources which were 
undetected at 95\,GHz, we used the upper limits in flux density to calculate a limiting 
value of $\alpha$.  
Figure \ref{fig.hist} shows the distribution of $\alpha_{20}^{95}$ 
for the two samples. 
Only two sources in the flux--limited sample (0227$-$3026 and 2158$-$3013) 
show a rising spectrum from 20--95\,GHz.   
Even in the inverted--spectrum sample (all of which were selected to 
have $\alpha_{8}^{20}>0$), only seven of the 37 sources have 
$\alpha_{20}^{95}>0$.  
This contrasts strongly with the 
spectral--index distribution of AT20G sources at 8--20\,GHz, 
where 40\% have rising spectra with $\alpha_{8}^{20}>0$ 
(Sadler et al.\ 2006).  As discussed by Lowe et al.\ (2007) this  
may be a selection effect, since the number of rising--spectrum 
sources in a sample selected at frequency $\nu$ will usually be greater 
when the spectral index is measured between $\nu$ and a lower frequency  
$\nu_1$ than when it is measured between $\nu$ and a higher frequency $\nu_2$. 

\begin{figure}
\begin{center}
\includegraphics{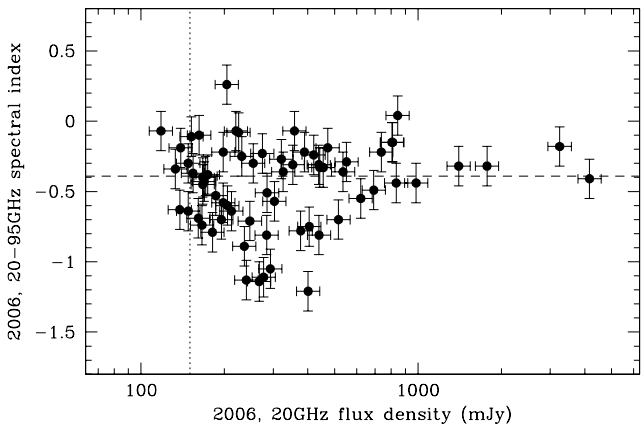} 
\vspace{6cm}
\caption{Plot of 20--95\,GHz spectral index $\alpha^{95}_{20}$ against 20\,GHz flux density 
for the flux--limited sample of sources in Table 4. The horizontal dashed line shows the 
median spectral index of $-$0.39, while the vertical dotted line shows the flux limit 
of 150\,mJy used to select the sample.  A few sources lie to the left of this line 
because their 20\,GHz flux density has decreased since the AT20G observations were 
made in 2004. } 
\label{fig.alpha}
\end{center}
\end{figure}

The median 20--95\,GHz spectral index for sources in our flux--limited sample 
is $-0.39$ and, as can be seen from Figure \ref{fig.alpha}, this 
does not change significantly with 20\,GHz flux density over the range 
probed by our ATCA observations (S$_{20}>150$\,mJy).  As we will show in \S5, 
this allows us to use the 20\,GHz extragalactic source counts measured from 
the AT20G survey, together with the observed distribution of $\alpha^{95}_{20}$, 
to derive the overall radio source counts at 95\,GHz. 

We note that our median 20--95\,GHz spectral index of $-0.39$ is much 
flatter than the value of $-0.89$ measured at 15--43\,GHz by Waldram et al.\ (2007). 
The Waldram et al.\ (2007) sources are much fainter than those in our 
sample, since most of them have 22\,GHz flux densities below the AT20G survey limit 
of 40\,mJy. The work of Waldram et al.\ (2007) is therefore complementary to our study, 
and suggests that the 20--95\,GHz spectral--index distribution of AT20G sources 
may steepen significantly at flux densities below 50--100\,mJy. 

\subsection{Properties of the 5--95\,GHz radio spectra} 
For the sources in this study, we have near--simultaneous data at 5, 8 and 20\,GHz 
from the AT20G survey as well as simultaneous data (at a different epoch) 
at 20 and 95\,GHz from our own ATCA observations.  
The mean and median values 
of the spectral index in each frequency interval are listed in Table \ref{tab.alpha}. 
Both samples show a general steepening of the radio spectrum with increasing frequency. 
A similar effect has been noted by Waldram et al.\ (2007) and Massardi et al.\ (2008). 

\begin{figure}
\begin{center}
\includegraphics{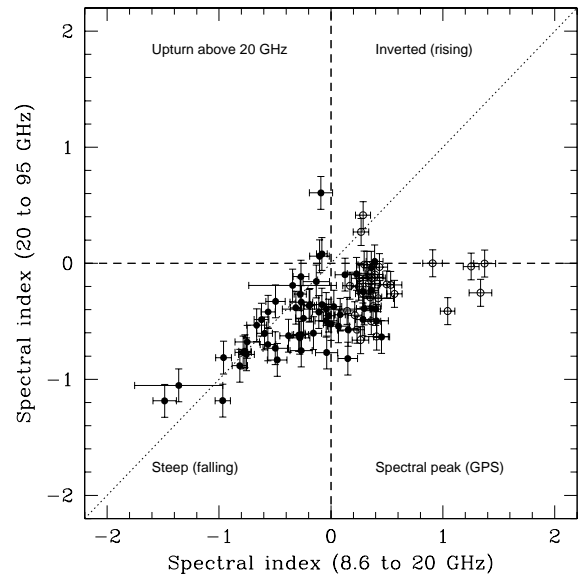} 
\vspace{8cm}
\caption{Two-colour radio spectral-index plot for the flux--limited 
(filled circles) and inverted--spectrum (open circles) AT20G samples.  
Only three sources still have a rising spectrum from 20--95\,GHz, 
and almost all the sources with inverted spectra at 8--20\,GHz have peaked 
and turned over by 95\,GHz.}
\label{fig.2col}
\end{center}
\end{figure}

Figure \ref{fig.2col} shows a radio two--colour diagram comparing the 
8--20 and 20--95\,GHz spectral indices.  As discussed by Sadler et al.\ (2006), 
this diagram provides a useful way of classifying objects which (like many of the 
AT20G sources) have curved, rather than power--law radio spectra. The diagonal 
dotted line in Figure \ref{fig.2col} shows the power--law relation. As noted 
earlier, inverted--spectrum objects (which account for 18\% of all AT20G sources at 
8--20\,GHz), are almost absent ($<2$\%) at 20--95\,GHz. 

\begin{table}
\begin{center}											
\begin{tabular}{ccccrl} 									
\hline												
 \multicolumn{1}{c}{Frequency} & Median & Mean & $\pm$ & n & Notes \\				
 \multicolumn{1}{c}{(GHz)} & \multicolumn{1}{c}{$\alpha$} & \multicolumn{1}{c}{$\alpha$} &  & \\ 
\hline												
 5--8   &  $-0.11$  & $-0.17$  & 0.06 & 62 & This paper \\					 
 8--20  &  $-0.24$  & $-0.25$  & 0.05 & 62 & This paper \\					
 20--95 &  $-0.39$  & $-0.45$  & 0.04 & 69 & This paper \\								
\hline												
\end{tabular}											
\caption{Mean and median values of the radio spectral--index distribution 
for our 2006 flux--limited sample calculated for different frequency ranges. 			
The three sources undetected at 95\,GHz have been removed from the calculation  		
of the mean value at 20--95\,GHz. }
\label{tab.alpha}										
\end{center}											
\end{table}											

\begin{table*}
\begin{minipage}{170mm} 
\begin{center}
\begin{tabular}{@{}llrrrccc@{}}
\hline											
\multicolumn{1}{c}{Sample} & N &  \multicolumn{1}{c}{Gal} & \multicolumn{1}{c}{QSO} & \multicolumn{1}{c}{Faint/blank} & 
\multicolumn{1}{c}{Median B$_{\rm J}$} & \multicolumn{1}{c}{N$_z$} & \multicolumn{1}{c}{Median $z$} \\				
    &&&&& \multicolumn{1}{c}{(mag)} & \\
\hline												
Inverted spectrum   & 34   &  7 ($21\pm8$\%) & 23 ($68\pm14$\%)   & 4 ($11\pm7$\%)    &  19.84 &  12 (35\%) & 0.56 \\					 
Flux--limited       & 69   & 11 ($16\pm5$\%) & 50 ($73\pm10$\%)   & 8 ($11\pm4$\%)    &  19.44 &  32 (46\%) & 1.11 \\								
\hline
\end{tabular}
\caption{Optical properties of the two samples (excluding sources for which no optical 
ID was attempted). N$_z$ is the number of sources in each sample which have a published optical 
redshift. Objects classed as `Faint/blank' have optical IDs fainter than B$_{\rm J}$=22.0. }
\label{tab.opt}
\end{center}
\end{minipage}
\end{table*}

\subsection{Optical properties of the 95\,GHz sources }
We used the Supercosmos catalogue (Hambly et al.\ 2001) to obtain optical 
identifications for the radio sources observed in this study.  We accepted 
an optical object as the correct ID if it was 
brighter than B$_{\rm J}=22$\,mag and lay within 2.5\,arcsec of the radio 
position, since Monte Carlo tests imply that at least 97\% of such objects 
are likely to be genuine associations (Sadler et al.\ 2006).  
Optical identifications were not attempted for three sources within 10\,degrees 
of the Galactic plane, and one source in the flux--limted sample (0320$-$3837) 
was so close to a bright foreground star that no optical identification was possible. 

Table \ref{tab.opt} summarizes the optical properties of the two samples.  While QSOs make up the majority 
of the optical IDs in both samples, the inverted--spectrum sample has a slightly higher fraction of galaxies 
and a median value of B$_{\rm J}$ 0.4 mag fainter than the flux--limited sample. 

Published redshifts are available for fewer than half the optical IDs, so the redshift distribution needs to be 
interpreted with caution. However, the lower median redshift for the inverted--spectrum sample is consistent 
with it containing a higher fraction of galaxies (which generally lie at $z<0.5$ if visible on DSS images) 
than the flux--limited sample. 

\subsection{ A correlation between 95\,GHz flux density and optical magnitude?} 
Owen \& Mufson (1977) found a correlation between optical magnitude and 
millimetre (90\,GHz) flux density for a sample of flat--spectrum QSOs, and noted  
that this was surprising because of the 20-25 year interval between their 
90\,GHz measurements and the sky survey plates from which the optical magnitudes 
were measured. They showed that the correlation was significant 
at the 99\% level at 90\,GHz, yet the 5\,GHz flux densities of the 
same sources were uncorrelated with optical magnitude. They interpreted this 
as evidence that the sources were optically thin at millimetre  wavelengths 
but optically thick at centimetre wavelengths, and that a correlation existed 
between the radiation mechanisms in the optical and millimetre regimes. 
 
\begin{figure}
\begin{center}
\includegraphics{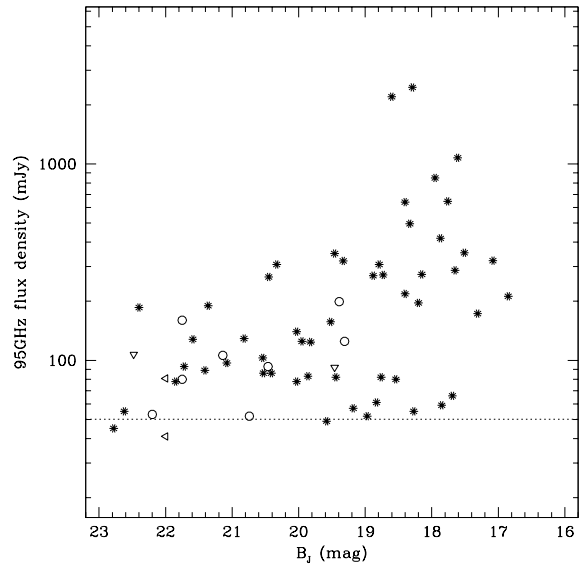} 
\vspace{8cm}
\caption{Plot of optical B$_{\rm J}$ magnitude from the SuperCOSMOS catalogue 
against 95\,GHz flux density for objects in our flux--limited sample.  
Sources identified with QSOs are shown by stars, and galaxies by open circles. 
Upper limits are indicated by open triangles, and the horizontal 
line at 50\,mJy shows the typical detection limit for our 2006 observations.  }
\label{fig.opt}
\end{center}
\end{figure}

Figure \ref{fig.opt} shows a plot of 95\,GHz flux density against optical 
B$_{\rm J}$ magnitude for our flux--limited sample, with separate symbols 
used for QSOs and galaxies. Blank fields are indicated by upper limits in 
B$_{\rm J}$.  We see a trend similar to that found by Owen \& Mufson (1977), 
in the sense that brighter 95\,GHz sources are associated with brighter 
optical objects.
The correlation for QSOs in Figure \ref{fig.opt} is statistically significant 
at a level of $>99.6$\% in a rank correlation test. 

One possible explanation is that many of the strongest  
95\,GHz sources are brightened by relativistic beaming at both optical 
and millimetre wavelengths, which would require that they are viewed at 
very small angles to the jet axis.  In this orientation, the millimetre 
continuum emission can be Doppler-boosted by factors of up to twenty 
(L\"ahteenm\"aki \& Valtaoja 1999) and the optical continuum brightened 
by at least one magnitude (Browne \& Wright 1985).  

In principle, optical spectroscopy can provide a test of the relativistic 
beaming model, since QSOs which are strongly beamed should have a lower line 
to continuum ratio than less beamed objects (Baker \& Hunstead 1995).  
This would require a more complete and uniform set of optical spectra than 
is currently available for our sample.

\subsection{Sources with extended low--frequency radio emission} 
SUMSS images at 843\,MHz are available for all our 95\,GHz sources, 
and 1.4\,GHz NVSS images for those north of declination $-40^\circ$. 
The SUMSS and NVSS images have similar resolution (45\,arcsec) and 
sensitivity (detection limits of a few mJy). 
Although all the sources in Tables 4 and 5 are unresolved 
at 20 and 95\,GHz, several of them show extended emission on scales of 
45--100\,arcsec in the low--frequency SUMSS and NVSS images (see Appendix A). 
From the NVSS/SUMSS images alone it is unclear whether the 
extended low-frequency emission arises only from the AT20G object, or 
includes a contribution from a nearby confusing source which is below 
the detection threshold at high frequencies. 

Interestingly all the sources with extended low--frequency emission 
are members of the inverted--spectrum sample.  Five of the 37 sources 
in the inverted--spectrum sample (14\%) have extended emission in the 
SUMSS images, whereas none of the 70 sources in the flux--limited 
sample do.  The reason for this difference is unclear at present.  
It may simply reflect  the lower median redshift 
of the inverted--spectrum sample (see Table \ref{tab.opt}), but could 
also hint at genuine physical differences between the inverted--spectrum 
sources (and/or their environment) and the overall high--frequency radio 
population. 

\section{ATCA 95\,GHz source counts and comparison with earlier predictions}
As noted earlier, there are several reasons for wanting to measure the  
surface density of extragalactic radio sources at 95\,GHz apart from the study 
of the intrinsic physical properties of these objects. 
In an ideal case one would perform an all-sky survey at 90--100\,GHz, 
but it is currently difficult to reach faint flux levels 
with such a survey.  
As noted in \S1, the primary beam of most radio interferometers and 
single--dish telescopes operating at mm-wavelength is tiny, making the 
mosaicing of any large area of sky an incredibly time-consuming 
task\footnote{For example, a blind survey carried at 100\,GHz with 
the same observational setup used for the AT20G survey at 20\,GHz 
(3 baselines; 8\,GHz analogue correlator and fast meridian scanning; 
see e.g. Sadler et al.\ 2006) on the same 1500\,deg$^2$ sky strip explored 
by the AT20G Pilot Survey ($-70^{\circ} < \delta < -60^{\circ}$; 
see Ricci et al.\ 2004) would require a month of observing time at the 
ATCA instead of the five days needed at 20\,GHz.  With a detection limit of 
$\sim300$\,mJy, this 100\,GHz survey would detect less than 
one source a day.}. 
Additional challenges include the high system temperatures at 90--100\,GHz 
and changing weather conditions which can make the sensitivity level of 
ground-based surveys uneven. 

The Planck satellite scheduled for launch in 2008  
will survey the whole sky with an estimated 5$\sigma$ point--source 
detection limit of $\sim500$\,mJy at 30\,GHz and $\sim280$\,mJy at 100\,GHz 
(L{\'o}pez--Caniego et al.\ 2006). These limits are set by confusion, and 
the rms noise per pixel is much lower (14\,mJy at 100\,GHz and
10.2\,mJy at 143\,GHz (Lamarre et al.\ 2003). As a result, there is a lot 
of astrophysical information in Planck maps below the confusion limit, and 
some of this can be extracted, e.g. using stacking techniques, by making use 
of the AT20G survey and follow-up observations at higher frequencies.

Since a blind survey of the whole sky to 
low flux levels at frequencies above 20\,GHz is currently impractical, 
we use a method first proposed by Kellermann (1964) to predict 
the extragalactic source counts at 95\,GHz, based on our knowledge of 
the observed 20\,GHz source counts from the AT20G survey and the 
observed 20--95\,GHz spectral--index distribution presented in \S4.5 
of this paper. 

\subsection{Methodology}
The technique developed by Kellermann (1964) is based on the idea that 
if one knows both the source counts at one frequency and the distribution 
of flux densities for a complete sub-sample at a second frequency, 
it is possible to predict the source counts at the second frequency.  

The method makes no assumption about the actual spectral shape 
between the two frequencies, but it is convenient to characterize 
the flux ratio by the spectral index for an assumed power law.  
It does assume that the distribution of flux ratios (i.e. the 
spectral--index distribution) correctly characterizes all the 
source populations present.  If the spectral--index distribution is a 
function of flux density, or if some class of sources becomes important 
at the second frequency but is not represented at the first frequency, 
the prediction will fail.  For this reason it is important to start with 
reliable source counts at a frequency as close as possible to the second 
frequency.  

The method we followed (outlined by Kellermann et al.\ 1968) first assumes  
that the ratio of the flux densities S$_1$ and S$_2$ at two frequencies 
$\nu_1$ and $\nu_2$ is represented by the spectral index $\alpha$ defined 
as  

\begin{equation} \label{eqn:spix}
S_2 = S_1 \times(\frac{\nu_2}{\nu_1})^{\alpha}
\end{equation}

\noindent
The observed spectral index distribution $P(\alpha)$ and the counts $n(S_1,\nu_1)$ 
at any given frequency $\nu_1$ above the flux density limit $S_1$ can then 
be used to predict the source counts $n(S_2,\nu_2)$ at another frequency $\nu_2$ 
above the flux density limit $S_2$ using the formula:

\begin{equation} \label{eqn:kgen}
n(S_2,\nu_2) = \int_{-\infty}^{+\infty} n[( \frac{\nu_2}{\nu_1} )^{-\alpha}\ S_2, \nu_1 ]\ P(\alpha)\ d\alpha
\end{equation}   
 
\noindent
provided that the $P(\alpha)$ is independent of flux density. 

For our flux--limited sample with S$_{20}>150$\,mJy and a median 20--95\,GHz 
spectral index of $-0.39$,  the equivalent flux--density limit at 95\,GHz is 82\,mJy. 

\subsection{Completeness of the 95\,GHz source populations} 
As noted above, the method we are using to predict the 95\,GHz source 
counts will give unreliable results if any class of sources 
which is present at 95\,GHz is not represented in our 20\,GHz source counts.  

In our case, any new population of sources appearing at 95\,GHz would 
have to have a very rapidly--rising spectrum between 20 and 95\,GHz 
in order to be weaker than the AT20G detection limit of 40\,mJy at 20\,GHz.
At the 1\,Jy level at 95\,GHz, this would require a population with a 
high--frequency radio spectral index $\alpha>+2.0$.  Even at the 100\,mJy 
level, a high--frequency spectral index $\alpha>+0.6$  would be required. 

No sources with such extreme 20--95\,GHz spectral indices are seen in 
our current 95\,GHz data set, apart from a small population of 
optically--thick thermal sources in our own Galaxy (which all lie at 
low Galactic latitude). As can be seen from Figures \ref{fig.hist} 
and \ref{fig.2col}, almost all AT20G sources with $\alpha>0$ at 
8--20\,GHz have lower 
flux densities at 95\,GHz than at 20\,GHz.  Bolton et al.\ (2004) also 
found no sources with $\alpha>+0.6$ at 15--43\,GHz in their sample 
of sources selected at 15\,GHz, with flux densities as low as 25\,mJy. 

At this stage, therefore, we see no evidence that that our 20\,GHz sample 
is missing any source population which contributes to the 95\,GHz source 
counts at levels above about 100\,mJy. 

\subsection{Differential radio--source counts at 95\,GHz} 
Since the 20\,GHz counts show some curvature, we made our final 
estimate by fitting the AT20G source counts with a double power 
law of the form: 

\begin{equation}
{\rm N(S)=2N_*/[(S/S_*)^\alpha + (S/S_*)^\beta]}
\end{equation}

\noindent
where the best--fit parameters for the two slopes $\alpha$ and $\beta$, 
the flux density of the break S$_*$ and number count at the break N$_*$ 
are given in Tables \ref{tab.fit} and \ref{tab.fit95}. 

\begin{table}
\begin{center}
\begin{tabular}{crl} 
\hline
Parameter & Fitted value & Units \\
\hline
$\alpha$ & $1.92^{+0.03}_{-0.02}$ & \\
$\beta$  & $2.78^{+0.09}_{-0.07}$ & \\
S$_*$    & $1.09^{+0.13}_{-0.18}$ & Jy \\
N$_*$    & $28.9^{+13.7}_{-6.8}$  & Jy\,sr$^{-1}$ \\
\hline
\end{tabular}
\caption{Best--fit parameters for the 20\,GHz source counts measured from 
the AT20G survey at declination $-90^\circ<\delta<-15^\circ$ (Ricci et 
al., in preparation). }
\label{tab.fit}
\end{center}
\end{table}

We then derived the differential source counts at 95\,GHz using the methods 
outlined in \S5.1. 
The results are shown in Figures \ref{fig.diff95} and \ref{fig.norm95}, \
and are valid for flux densities above 80\,mJy at 95\,GHz. 

\begin{table}
\begin{center}
\begin{tabular}{crl} 
\hline
Parameter & Fitted value & Units \\
\hline
$\alpha$ & 1.92 & \\
$\beta$  & 2.77 & \\
S$_*$    & 0.80 & Jy \\
N$_*$    & 19.45 & Jy\,sr$^{-1}$ \\
\hline
\end{tabular}
\caption{Parameterization of our predicted 95\,GHz source counts  
plotted in Figure \ref{fig.diff95}.  The fit was done in the same way as for 
the 20\,GHz counts in Table \ref{tab.fit}.  }
\label{tab.fit95}
\end{center}
\end{table}

\subsection{Comparison with previously--published values}  
\begin{figure}
\begin{center}
\includegraphics{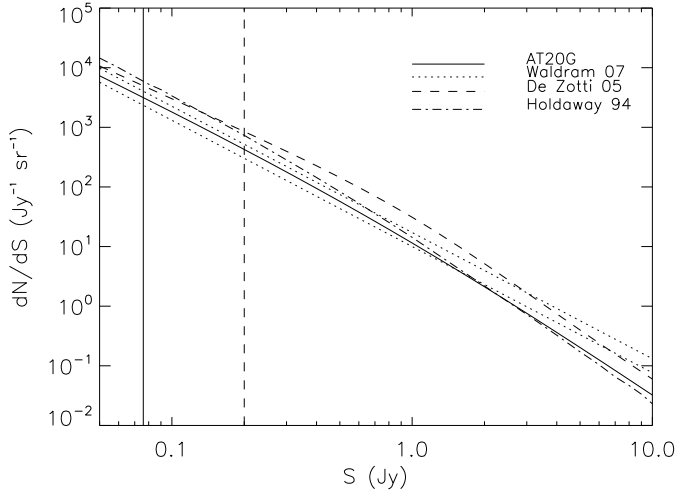} 
\vspace{8cm}
\caption{Differential source--count predictions from AT20G 20\,GHz counts 
scaled to 95 GHz (this paper); De Zotti et al. (2005) model counts at 94 GHz; 
Waldram et al. (2007) at 94 GHz and Holdaway et al. (1994) at 90 GHz. 
The vertical solid line represents the lower bound of validity for our counts 
(82\,mJy), and the vertical dashed line is the upper bound of validity for 
the Waldram et al.\ counts (200\,mJy). }
\label{fig.diff95}
\end{center}
\end{figure}

\begin{figure}
\begin{center}
\includegraphics{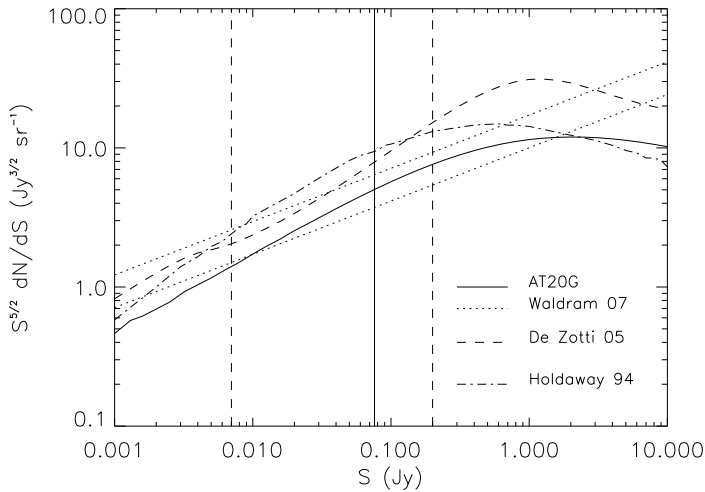} 
\vspace{8cm}
\caption{Differential 95\,GHz source counts normalized to the Euclidean case 
for the same four estimates shown in Figure \ref{fig.diff95}.   
The flux--density scale now extends down to 1\,mJy. As before, the vertical 
solid line represents the lower bound of validity for our counts. The two 
vertical dashed lines show the lower and upper bounds of validity for the 
Waldram et al.\ (2007) predictions. 
}
\label{fig.norm95}
\end{center}
\end{figure}

Figures \ref{fig.diff95} and \ref{fig.norm95} also compare 
our 95\,GHz differential source counts with previous predictions by 
Holdaway et al.\ (1994), De Zotti et al. (2005) and Waldram et al.\ (2007). 

Holdaway et al.\ (1994) observed a sample of 367 flat--spectrum radio 
sources at 90\,GHz with the NRAO 12\,m telescope in 1993. 
Just under 80\% of their sample was detected above a 3$\sigma$ 
limit of 90\,mJy.  
Holdaway et al.\ (1994) then used VLA 8.4\,GHz measurements 
of the same objects (Patnaik et al.\ 1992) to calculate the 8--90\,GHz 
spectral index distribution for their sample, and combined this with 
the 5\,GHz source counts measured by Condon (1984) to estimate the 
source counts at 90\,GHz. 

The 8 and 90\,GHz observations were separated in time by almost 
four years (the data were taken in 1990 Feb and 1993 Nov respectively), 
and so the measured 8--90\,GHz spectral index distribution is likely 
to be broadened by source variability over the intervening period of 
almost four years. The Holdaway et al.\ (1994) analysis also depends 
critically on the assumed ratio of core to total flux density at 5\,GHz, 
since the 5\,GHz source counts measure total flux density 
but their 8--90\,GHz data essentially measures the spectral index 
of a compact core.  

Their predicted source counts from Holdaway et al.\ (1994) are valid for 
the flux--density range S$_{90}>0.1$\,Jy, set by the detection limit of 
the high--frequency survey they used.  

Radio source counts at 94\,GHz have been estimated using 
the evolutionary models described by De Zotti et al.\ (2005), which take into 
account the epoch--dependent radio luminosity functions of the various source 
populations which contribute to the high--frequency counts. 
The 94\,GHz counts are derived by making the additional 
assumption that all flat--spectrum sources have a simple power--law spectrum 
with $\alpha=-0.1$ above 20\,GHz (De Zotti et al.\ 2005). 

The De Zotti et al.\ (2005) model counts at 20 and 30\,GHz are calculated over 
a wide range in flux density (0.01\,mJy to 10\,Jy), though De Zotti et 
al.\ (2005) note that there are large uncertainties below $\sim$0.1\,mJy.  
Both Sunyaev--Zeldovich signals 
and free--free emission from proto-galactic plasma are likely to contribute 
significantly at sub--mJy flux levels, but there are currently almost no 
observational data available to constrain the models. 

Waldram et al.\ (2007) have recently estimated the source counts at 90\,GHz 
based on the 15\,GHz source counts from their 9C survey (Waldram et al.\ 2003)
combined with follow--up 22 and 43\,GHz VLA observations of a flux--limited 
sample of 121 sources selected at 15\,GHz. 
The Waldram et al.\ (2007) source--count predictions are valid for 
the flux--density range between 10 and 200\,mJy, since their 15\,GHz survey 
area contains very few strong sources.  

Our derived 95\,GHz source counts are in good agreement with the counts 
published by Waldram et al.\ (2007) in the 
flux density range where both data sets are valid (80--200\,mJy; see 
Figure \ref{fig.norm95}). This is encouraging, as the two derivations 
are based on completely independent data sets. 

In contrast, our 95\,GHz counts are significantly lower than the values derived by 
Holdaway et al.\ (1994) for all but the brightest sources, and are also lower 
than the De Zotti values at all flux densities. In the case of the De Zotti 
counts, the discrepancy almost certainly arises because De Zotti assumed a power--law 
spectral index of $-0.1$ at 20--94\,GHz, whereas our 95\,GHz data show that 
the median spectral index of AT20G sources in this frequency range is significantly 
steeper ($\sim-0.4$, see Table \ref{tab.alpha}). 

For the Holdaway et al.\ (1994) counts, the situation is more complex.  As they note 
in their paper, the exact form of the inverted--spectrum tail of their 8--90\,GHz 
spectral--index distribution (which is poorly defined because of the small number of 
sources observed in this region) is very important because it allows a small fraction 
of the numerous weak 5\,GHz sources to be boosted up to very high flux densities at 90\,GHz. 
Holdaway et al.\ (1994) do not discuss the effects of variability over the several--year 
interval between their 8 and 90\,GHz datasets, but this could also scatter variable objects 
into the tail of their spectral--index distribution. Since our results and those of 
Waldram et al.\ (2007) agree well and are also based on higher--frequency (15--20\,GHz) 
source--count data, we conclude that Holdaway et al.\ (1994) have over-estimated 
the surface density of 95\,GHz sources by up to a factor of two. 

\subsection{Integrated source counts at 95\,GHz}
As a consistency check, we can use the differential source counts derived 
in the previous section to estimate the number of extragalactic radio sources brighter 
than a certain S$_{\rm lim}$ over the whole sky at 90--95~GHz:

\begin{equation} \label{eqn:intcnt}
N(S_{90} > S_{\rm lim}) = 4\pi\ \int_{S_{\rm lim}}^{+\infty}\ n(S)\ dS 
\end{equation}    

Our results for S$_{\rm lim}=0.1$ and 1.0 Jy are reported in 
Table~\ref{tab.intcnt} together with predictions from De Zotti 
et al.\ (2005) and Holdaway et al.\ (1994).  Since the 
Waldram et al.\ (2007) counts are only valid for sources below 200\,mJy, 
integrated source counts cannot be derived from their study. 

\begin{table}
\begin{center}
\begin{tabular}{lcll}
\hline
   &    $\nu$ & \multicolumn{2}{c}{Integrated counts (all--sky)} \\
Source        & (GHz) & $>$0.1\,Jy & $>$1\,Jy  \\
\hline
This paper              &  95 &  $2310\pm1132$  & $120\pm64$   \\
De Zotti et al.\ (2005) &  94 &  4457  & 288    \\  
Holdaway et al.\ (1994) &  90 &  4400  & 178    \\
\hline
\end{tabular}
\caption{Comparison of the all-sky integrated source counts above a limiting 
flux density S$_{\rm lim}$ at the frequency $\nu$ between our predictions and 
the De Zotti et al.\ (2005) model and Holdaway at et. (1994) estimates.  
Our quoted errors have been derived from a Monte Carlo simulation based 
on the 20\,GHz source counts and 20--95\,GHz spectral--index distribution, 
and are dominated by the small--number statistics in the spectral--index 
distribution.   }
\label{tab.intcnt}
\end{center}
\end{table}

Our derived source numbers are significantly lower than those derived by 
both De Zotti et al.\ (2005) and Holdaway (1994).  This is a reflection 
of the lower differential source counts found in our study, as discussed in the 
previous section.  The actual number of sources stronger than 1\,Jy over 
the whole sky at 90--95\,GHz is still poorly determined, but the WMAP 
catalogue (Hinshaw et al.\ 2007) contains 106 sources which lie above 
this flux level at 94\,GHz. The ATCA calibrator catalogue at declinations 
south of 0$^\circ$ contains 60 extragalactic sources stronger than 1\,Jy 
in the 86--100\,GHz band, implying roughly 120 such sources over the 
whole sky.  Our derived value in Table \ref{tab.intcnt} 
therefore appears reasonably consistent, within the errors, 
with what is observed. 

\subsection{Confusion noise in CMB surveys at 95\,GHz}
The 95~GHz source counts can also be used to compute the confusion noise 
$\sigma_{\rm conf}^2$ and the angular power spectrum $\delta T_{l}$ of the 
temperature fluctuations due to discrete point-like extragalactic radio sources, 
seen as a small angular scale contamination to the Cosmic Microwave Background 
anisotropies. This study is complementary to the work described in the previous 
section, because once the brightest sources in the 90--100\,GHz sky have been 
identified and removed from the CMB maps, the remaining fainter sources act 
as a confusing background. 
The formalism to derive $\sigma^2$ and $\delta T_{l}$ is based 
on Cleary et al.\ (2005) and briefly described below.       

Once a limiting flux density (S$_{lim}$) for a survey at a certain observing 
frequency has been determined, the contribution to the temperature fluctuations
from faint sources below that flux density limit has two components: one 
$C_{\rm src}^{\rm poisson}$ due to the Poisson--distributed sources and 
a second term due to clustering $C_{\rm src}^{\rm clustering}$:

\begin{equation} \label{eqn:aps}
T_{\rm CMB}^2\ C_{\rm src} = (\frac{\partial B_{\nu}}{\partial T})^{-2}\ (C_{\rm src}^{\rm poisson} + C_{\rm src}^{\rm clustering})
\end{equation}

$C_{\rm src}$ is the radio source angular power spectrum and 

\begin{equation} \label{eqn:fct}
\frac{\partial B_{\nu}}{\partial T} = \frac{2k}{c^2}\ (\frac{k\ T_{\rm CMB}}{h})^2\ \frac{x^4\ \exp^{x}}{(\exp^{x} - 1)^2} 
\end{equation}    

is the conversion factor between fluctuations in the background intensity and 
in antenna temperature. k here is the Boltzmann's constant and 
$x\equiv h\nu/kT_{\rm CMB}$, where $T_{\rm CMB}=2.726$~K is the CMB mean 
temperature. The term $C_{\rm src}^{\rm poisson}$ given by the formula: 

\begin{equation} \label{eqn:cnoise}
C_{src}^{\rm poisson}\equiv\sigma_{\rm conf}^2 = \int_{0}^{S_{\rm lim}}\ S^2\ n(S)\ dS
\end{equation}       

is also known as {\it confusion noise} for Poisson--distributed sources. 
This quantity is easy to estimate starting from experimental differential 
source counts $n(S)$.   

The term for source clustering (Scott \& White 1999) in equation 
(\ref{eqn:aps}) is given by:

\begin{equation} \label{eqn:clust}   
C_{\rm src}^{\rm clustering} = \omega_{l}\ (\int_{0}^{S_{\rm lim}}\ S\ n(S)\ dS)^2
\end{equation}

where $\omega_{l}$ is the Legendre transform of the angular two-point 
correlation function of radio sources $\omega(\theta)$. Lacking any clear 
estimate of the clustering properties of extragalactic radio sources at the 
frequency of 90~GHz, we assume for now that this term is negligible.

In order to compare our prediction with others we conventionally expressed 
the point-source angular power spectrum in terms of the Legendre multiple 
coefficient $l$:  

\begin{equation} \label{eqn:dtl2}
\Delta T_{\rm src}^2 = T_{\rm CMB}^2 \frac{l(l+1)\ C_{\rm src}}{2\pi} \approx const. \times (\frac{l}{100})^2
\end{equation}        

so that the $\delta T_{l}$ is simply given by:

\begin{equation} \label{eqn:dtl}
\delta T_{l} = (\Delta T_{\rm src}^2)^{1/2} \approx A \times \frac{l}{100}
\end{equation}

The normalization coefficient $A$ of the angular power spectrum $\delta T_{l}$
obtained from our 95\,GHz predicted counts is shown in 
Table~\ref{tab.apscomp} together with the model by Toffolatti et al.\ (1998), 
the model by De Zotti et al.\ (2005) which includes the contribution of 
flat-spectrum radio quasars, BL Lacs and steep-spectrum radio galaxies and 
the estimate by Giommi et al. (2006) for the blazar population only. In our 
computation of $\delta T_{l}$ we assumed that the 95\,GHz counts 
keep the same functional form for $S_{\rm 95} < 82$ mJy, which is the  
effective 95\,GHz limit for our target sample.

\begin{table}
\begin{center}
\begin{tabular}{lcc}
\hline
Author & $\nu$    &  A \\
       & (GHz)    & ($\mu$K) \\
\hline
This paper      & 95    & 0.78        \\
De Zotti et al.\ (2005)   & 94    & 1.12        \\  
Toffolatti et al.\ 1998 & 100   & 1.0         \\
Giommi et al.\ 2006     & 94    & 1.3         \\   
\hline
\end{tabular}
\caption{Comparison between the normalization coefficient $A$ of temperature 
fluctuation angular power spectrum due to the background of extragalactic 
discrete point-like sources fainter then 1 Jy at the observing frequency $\nu$.}
\label{tab.apscomp}
\end{center}
\end{table}

\section{Summary and future work} \label{sec:concl}
In this study, we have shown that the ATCA can measure robust 95\,GHz 
(3\,mm) flux densities 
for relatively faint continuum sources in a few minutes of integration 
time, provided that careful attention is given to calibrating the telescope pointing. 
Our ATCA measurements at 20 and 95\,GHz appear to be on the same flux scale as the WMAP 
catalogue (Hinshaw et al.\ 2007) to within the errors. 
The arrival in 2008 of the new ATCA broadband correlator, which has a 2\,GHz bandwidth 
compared to the current value of 128\,MHz, will significantly increase the continuum 
sensitivity at 95\,GHz and should make it possible to measure 95\,GHz sources down to 
$\sim10$\,mJy with integration times similar to those used here. 

We have made a new estimate of the 95\,GHz radio source counts for sources stronger 
than 80\,mJy, and show that some previous studies have over-estimated the number 
of these sources by up to a factor of two.  In particular, the confusion level 
in the Planck beam at 95\,GHz is likely to be significantly lower than previously 
estimated.  

Fewer than half the sources in our 95\,GHz sample currently have measured 
redshifts, and obtaining good-quality optical spectra for the whole sample 
would be valuable both to study their rest--frame radio spectra and to test 
whether the brightest 95\,GHz sources are relativistically beamed in the 
optical, as suggested in \S4.8.  

We plan to make new measurements of a complete sample of sources from the AT20G Bright 
Source Sample (Massardi et al.\ 2008) at 20, 40 and 95\,GHz in late 2007, using the new 
ATCA 7\,mm receiver system.  These observations, which will also include polarization 
measurements, should provide further insights into the properties of the strongest 
95\,GHz sources. 

\section{Acknowledgments}
We thank the staff at the Australia Telescope Compact Array for their 
support during our observing runs, and the AT20G and 6dFGS teams for 
making their data available to us in advance of publication.  
We also acknowledge the support of the Australian Research Council 
through the award of an ARC Australian Professorial Fellowship to EMS 
and a Federation Fellowship to RDE. 

The Australia Telescope Compact Array is part of the Australia Telescope, which is funded 
by the Commonwealth of Australia for operation as a National Facility managed by CSIRO.  
This research has made use of the NASA/IPAC Extragalactic Database (NED) which is 
operated by the Jet Propulsion Laboratory, California Institute of Technology, 
under contract with the National Aeronautics and Space Administration.
This research has also made use of data obtained from the SuperCOSMOS Science Archive, 
prepared and hosted by the Wide Field Astronomy Unit, Institute for Astronomy, 
University of Edinburgh, which is funded by the UK Particle Physics and Astronomy 
Research Council.

We thank the referee, Prof. Ian Browne, for a prompt and very helpful report.

\appendix 
\section{Notes on individual sources} 

\indent
{\bf 0058$-$3234:} The 6dFGS spectrum (Jones et al.\ 2004) shows a bright 
featureless continuum and no redshift could be measured. 

{\bf 0111$-$4749: } The optical ID is a 2dFGRS (Colless et al.\ 2001) 
galaxy at $z$=0.154 whose spectrum shows strong, narrow emission lines.  
This is one of the few AT20G sources in our sample which was 
undetected at 95\,GHz.  The radio spectrum rises steeply from 0.8 to 20\,GHz, 
and so must peak somewhere above 20\,GHz. The 843\,MHZ SUMSS images shows 
extended emission with a deconvolved largest angular size of 83\,arcsec 
at PA 36$^\circ$.

{\bf 0136$-$4044: }  This appears to be the same object as PMNJ\,0136$-$4044, 
though the catalogued PMN position is significantly offset from the more 
accurate AT20G radio position. There is a faint (stellar) optical ID 
at the AT20G position.  

{\bf 0142$-$4206: } This source appears extended in the 843\,MHz SUMSS
image (deconvolved major axis 59\,arcsec at PA 25$^\circ$).  

{\bf 0245$-$4459:} The 6dFGS spectrum shows strong, broad Balmer emission lines 
typical of a Seyfert 1 galaxy or QSO. 

{\bf 0320$-$3837: } This source is so close to a bright star that no optical is 
ID possible. 

{\bf 0428$-$4357: } Blank field on the blue DSS image.  No known optical ID. 

{\bf 0433$-$4502: } One of the few sources in our sample which shows an inverted radio 
spectrum at all frequencies from 0.8 to 95\,GHz.  The DSS optical ID is a 19th magnitude 
galaxy whose redshift is currently unknown. 

{\bf 0440$-$4731: } This source is well below the limit of the SUMSS catalogue at 843\,MHz, 
but examination of the original SUMSS image shows a marginal (3\,$\sigma$) detection of a 
2.5\,mJy source at the AT20G position.  There is no optical ID, and this is one of the few 
AT20G sources which does not appear in any previous optical or radio catalogue. 

{\bf 0523$-$2955: } This source is extended (or perhaps confused) in the SUMSS 
image (deconvolved major axis 47\,arcsec at PA\,141$^\circ$).  The NVSS image also 
shows extended emission with largest angular size of 56\,arcsec. 

{\bf 0608$-$3041: } This source is clearly extended north--south in the 
SUMSS image (deconvolved major axis 92\,arcsec at PA\,1$^\circ$), and an 
extended structure is also clearly seen in the AT20G 5\,GHz image 
even though the image quality is poor in this snapshot observation.  
The 1.4\,GHz NVSS resolves the source into two components separated 
by about 70\,arcse.  At 20\,GHz the source is unresolved. 
The 6dFGS spectrum shows strong, broad Balmer emission lines 
typical of a Seyfert 1 galaxy or QSO. At z=0.2237, this is a relatively nearby
object. 

{\bf 1117$-$4837: } This source is extended in the SUMSS image (deconvolved major axis 
47\,arcsec at PA\,75$^\circ$). 

{\bf 1427$-$3305: } The 6dFGS spectrum shows a featureless continuum, and no redshift 
could be measured. The optical counterpart is stellar on the blue DSS image, but is 
fuzzy (and classified as a galaxy) on the red image, suggesting that this may be 
a relatively nearby BL Lac object.

{\bf 1454$-$3747: } The 6dFGS spectrum shows strong, broad Balmer emission lines 
typical of a Seyfert 1 galaxy or QSO.

{\bf 2031$-$3647: } The 6dFGS spectrum shows emission lines of CIII] and MgII. 

{\bf 2113$-$3838: } The 6dFGS spectrum shows a strong blue continuum with a 
single broad emission line of MgII. 

{\bf 2113$-$3838: } The 6dFGS spectrum shows strong, broad emission lines 
of H$\gamma$ and MgII. 

\end{document}